\documentclass[proof]{WileyASNA-v1}

\usepackage{amstext}
\articletype{Original article}

\received{DD MM, 2025}
\revised{DD MM, 2025}
\accepted{DD MM, 2025}

\begin{document}

\title{Spatial extent of the magnetic influence of an active region on the solar atmosphere}

\author[1]{Gottfried Mann*}
\author[1]{Fr\'ed\'eric Schuller}
\authormark{Mann \& Schuller -- Magnetic influence of an active region}

\address[1]{\orgname{Leibniz-Institut f\"ur Astrophysik Potsdam}, \orgaddress{An der Sternwarte 16, 
    D-14482 Potsdam}, \country{Germany}}
    
\corres{*G.~Mann, \email{GMann@aip.de}}

\abstract[Abstract]
{Solar activity can be witnessed in the form of sunspots and active regions, where the magnetic field is enhanced by up to a factor 1000 as compared to that of the quiet Sun. In addition, solar activity manifests itself in terms of flares, jets, coronal mass ejections, and production of highly energetic particles. All these processes are governed by the solar magnetic field.
Here we study the spatial reach of the influence of the magnetic field of active regions on the photosphere and in the solar corona. An active region is modelled by a magnetic dipole located under the photosphere. This simplified description  allows us to study the spatial influence of an active region in the solar atmosphere in a rough but easy way.
We find that the area of influence of the magnetic field of an active region on the solar atmosphere increases with, both, the maximum strength of the magnetic field in the active region and the depth of the dipole under the photosphere. For a typical active region, the magnetic field can be neglected for distances beyond ca. 700~Mm on the photosphere and two solar radii in the solar corona.} 


\maketitle

%
\section{\label{sec-introduction}Introduction}
The Sun is an active star. Sunspots are an obvious manifestation of this activity.
\citet{Hale1908} observed strong magnetic fields up to 4000~G within sunspots. The occurrence of these spots on the disk of the Sun follows the well-known 11-year cycle.
A very strong magnetic field of 5500~G was even measured in the active region NOAA 12673 \citep{Wang2018,Lozitsky2022}.
This active region was the source of several X-GOES-class
flares as the solar event on September 10th, 2017 \citep{Veronig2018}. It was an X8.2 flare and the second largest one in solar cycle 24.
The activity of the Sun also manifests itself in the form of short-lived phenomena such as solar flares (see \citet{Priest1982}, \citet{Stix2004}, and \citet{Aschwanden2005} as text books and \citet{Benz2017} as a review),
solar radio bursts \citep{Wild1950}, and coronal mass ejections \citep[CMEs; see e.g.][for a review]{Temmer2021}.
A flare occurs as a sudden enhancement of the local emission of electromagnetic waves covering the whole spectrum from the radio up to the $\gamma-$ray range with durations of typically several hours
(see \citet{Aschwanden2005} as a textbook and \citet{Benz2017, Klein2021} for reviews).
During flares a large amount of stored magnetic energy is suddenly released and transferred into the local heating of the coronal plasma, mass motions (as e.g.\ jets), and the acceleration of highly energetic particles, a phenomenon usually referred to as "solar energetic particle events"
\citep[SEP; see e.g.][as reviews]{Lin1974,Klein2001,Zharkova2011,Mann+2018}.

Solar activity has an influence on our Earth's environment and, hence, on our technical civilisation \citep[see e.g.][as a review]{Temmer2021}.
That is usually called {\it space weather}.
All these phenomena are governed by the local and global magnetic field of the Sun and are rooted in magnetic active regions in the photosphere \citep[see][as a review]{Toriumi2019}.
Therefore, the knowledge of the structure of the magnetic field on the photosphere and in the solar atmosphere is of high interest for understanding the solar activity.

Active regions of the Sun appear in different formations (see \citet{Priest1982} and \citet{Stix2004} as textbooks).
On the one hand, some sunspots show a nearly unipolar high magnetic field. They are diffusely surrounded by small areas of oppositely polarised magnetic fields. The magnetic field in these areas is much weaker than in the leading spot. On the other hand, there are pronounced bi-polar active regions where the oppositely polarised areas have nearly the same magnetic field strength. In reality, the precise morphology of an active region is much more complex.

The aim of this paper is to study the spatial scale where the influence of the magnetic field of an active region dominates that of the quiet Sun in the solar atmosphere.
For simplicity, the magnetic field structure of an active region is modelled by a magnetic dipole, which is completely characterised by only three parameters, namely the maximum of the
magnetic field strength, the spatial size of the active region, and the angle of the
magnetic dipole with respect to the vertical axis.
These parameters can be determined by photospheric measurements. Our simplified model then
allows us to estimate the magnetic field strength in the neighbourhood of an active region in an easy and quick manner.

Furthermore, this method allows to estimate the spatial reach of the influence of the magnetic field of an active region in its environment.
Or in other words, what is the spatial scale at which the magnetic field strength of an active region decreases down to that of the quiet Sun level.
Outside active regions, the magnetic field strength is that of the (so-called) quiet Sun, which is typically a few Gauss \citep{Bellot2019}.
For instance, investigations of global EUV waves on the disk of the Sun revealed that the quiet Sun magnetic field is about 3.2~G \citep{Mann+1999,Klassen2000,Mann-Veronig2023}.

Modern numerical tools exist in order to extrapolate the magnetic field that is measured at the photospheric level into the corona \citep[see][as a review]{Wiegelmann2021}.
For instance, one well-known method is the potential-field-source-surface (PFSS) model \citep{Schrijver2003}.
It allows to determine the topology of the magnetic field in the corona in a highly accurate way 
\citep[see e.~g.][]{Erdelyi2022}.
But this requires precisely measured photospheric magnetic field maps as inputs for the extrapolation, as well as the implementation of computing-intensive algorithms.

The influence of the magnetic field of an active region on its environment in the solar atmosphere is discussed in Sect.~\ref{sec-influence}. We analyse the spatial scale of the influence of an active region as a function of the maximum magnetic field and the spatial size of the active region (or the depth of the associated dipole under the photosphere). The results of the paper are briefly summarised in Sect.~\ref{sec-summary}.

\section{Influence of the magnetic field of an active region on its environment}\label{sec-influence}
The aim of the present paper is to characterise the influence of the magnetic field
of an active region on its environment, and to evaluate the spatial scale where
the strength of this magnetic field decreases down to that of the quiet Sun.
For this purpose, an active region is modelled by a magnetic dipole for simplification.
For the sake of completeness, the structure of the magnetic field of a dipole is fully described in Appendix~A.

\begin{figure}[t]
  \centering
   \includegraphics[width=0.40\textwidth]{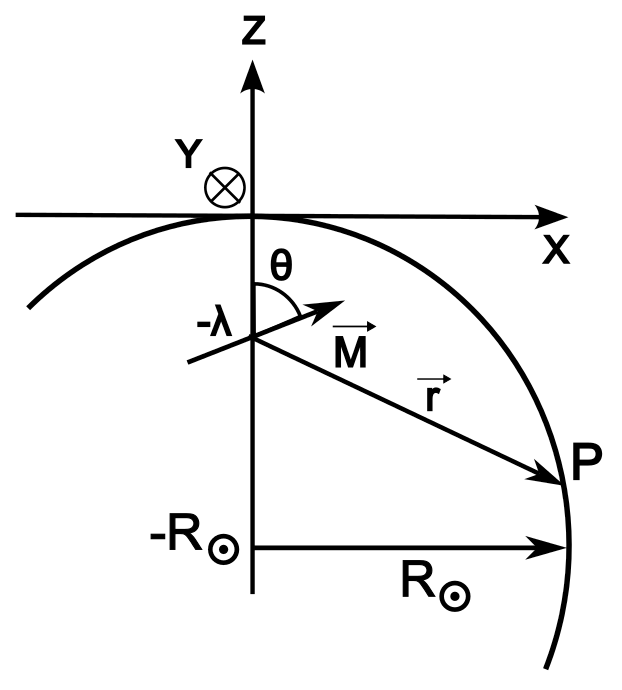}
   \caption{Sketch of the framework employed in this paper.
   \label{fig-sketch}}   
\end{figure}

The framework employed in this paper (see Fig.~\ref{fig-sketch}) is chosen in the following way: 
The z-axis is vertically directed away from the Sun. The centre of the Sun is located at the point
P$_{\odot}$ = (0,0,-$R_{\odot}$) (with $R_{\odot}$ being the radius of the Sun).
Thus, the surface (or photospheric level) of the Sun is described by:
\begin{equation}
\label{eq-photosphere}
R_{\odot}^{2} = x^{2} + y^{2} + (z+R_{\odot})^{2}
\end{equation}

The magnetic dipole is located in the x-z plane at a depth z = -$\lambda$ under the photosphere
and takes an angle $\vartheta$ with respect to the z-axis. Then, at a point P = $(x,y,z)$,
the vector $\vec r$ is given by $\vec r = (x,y,z+\lambda)$.
According to our chosen framework, the point P$_{centre}$ = (0, 0, 0) can be considered as the centre of the active region (see Fig.~\ref{fig-sketch}).
Then, the x-y-plane at z = 0 describes the photospheric level
in the neighbourhood of P$_{centre}$. The neutral line of an active region is usually defined by $B_{z}(x, y, z=0) =0$ at the photospheric level, leading to:
\begin{equation}
\label{eq-neutral-line}
\left ( x - \frac{3}{2} \cdot \lambda \cdot \tan \vartheta \right )^{2} + y^{2} = \frac{\lambda^{2}}{4} \cdot (8+9\tan^{2}\vartheta)
\end{equation}
by means of Eq.~(\ref{eq-B_z}). Eq.~(\ref{eq-neutral-line}) represents a circle with the origin at $x = 3\lambda\tan\vartheta/2$ and the radius $\lambda (8+9\tan^{2}\vartheta)^{1/2}/2$. In the special case of $\vartheta = 90^{\circ}$ the circle degenerates into a straight line along the y-axis at $x = 0$.

As shown in Appendix~A, the magnetic field of a dipole can
be completely determined by three parameters, namely by the 
magnetic field strength $B_{0}$ at the centre of the dipole at the 
photospheric level (i.~e. at P$_{centre}$), the depth $\lambda$ at which the dipole is 
located under the photosphere, and the angle $\vartheta$ between
the magnetic moment $\vec M$ of the dipole and the z-axis (see
Fig.~\ref{fig-sketch}). The special cases of a horizontal dipole ($\vartheta = 90^{\circ}$) and a vertical dipole ($\vartheta = 0^{\circ}$)
are discussed in detail in Appendix~A.
For illustration, the behaviour of the z-components of the magnetic field along the $x$-axis is presented for both cases in Figures~\ref{fig-B-hori} and \ref{fig-B-verti}. 

In the following, we consider the extreme cases of a dipole horizontally and vertically directed with respect to the solar surface.
In the case of the horizontal dipole, the z-component of the magnetic field has a minimum and a maximum with the same magnetic field strength (see Fig.~\ref{fig-B-hori}). Hence, a horizontal dipole describes roughly a bi-polar active region with a pronounced straight neutral line. 
The distance between both extrema is equal to the parameter $\lambda$, that is the depth of the magnetic dipole below the photosphere.
 
In the case of the vertical dipole, the z-component of the magnetic field has a dominant maximum at $\eta = 0$ and two weak minima at $\eta = \pm 2.0$, with $\eta = x/\lambda$ (see Fig.~\ref{fig-B-verti}),
with $|b_{min}/b_{max}|$ = 0.0177. Hence, the vertical dipole models roughly an uni-polar active region with a pronounced area of a high magnetic field surrounded by an oppositely directed weak magnetic field and a circular neutral line. 

Thus, the magnetic field is completely described by the parameters $B_{0}$ and $\lambda$ in both cases. They can be derived from photospheric measurements as the maximum magnetic field strength of the active region and the separation between the extrema in both polarities.
These measurements can be obtained with existing instruments, such as the Heliospheric Magnetic Imager \citep[HMI;][]{Scherrer2012,Schou2012}, and
the High Resolution Telescope in the Polarimetric and Heliospheric Imager
\citep[PHI/HRT;][]{Solanki2020,Sinjan2022} onboard the spacecraft {\it Solar Dynamics Observatory}
and {\it Solar Orbiter}, respectively.

In the following, we will use the parameters $B_{0}$ = 500~G, 1000~G, 2000~G, and 4000~G, and $\lambda$ = 15~Mm, 25~Mm, and 50~Mm. This choice covers a broad range of what is usually observed in typical active regions.
This is also consistent with observations of solar flares in hard X-rays, 
where non-thermal emission often shows double foot-point sources anchored
in the chromosphere. These sources are located in the vicinity of the flare ribbons,
which are located near the maxima and minima of the magnetic field at 
both sides of the neutral line.
\citet{Warmuth2013} studied a sample of 24 flares covering all GOES-classes using RHESSI data \citep{Lin2002}.
Their statistical analysis showed that the double sources 
(and, hence, the flare ribbons) have typical separations of 20-30~Mm.

In Sect.~\ref{sec-behaviour-surface} the behaviour of the magnetic field of the dipole on the solar surface or photospheric level is discussed. Subsequently, the vertical (or radial) behaviour of the magnetic field in the solar corona is studied in Sect.~\ref{sec-behaviour-radial}.
The results are briefly summarized in Sect.~\ref{sec-summary}.

\subsection{\label{sec-behaviour-surface}Behaviour of the magnetic field on the solar surface}
All computations in this section are done for $y = 0$.
The surface (or photospheric level) of the Sun is described by Eq.~(\ref{eq-photosphere}). Using $x' = x/R_{\odot}$, $z' = z/R_{\odot}$, and $\lambda' = \lambda/R_{\odot}$, this is equivalent to:
\begin{equation}
z' = - 1 + \sqrt{1-x'^{2}}
\end{equation}
Then, the length $s$ of the path from the point P$_{centre}$ = (0,0,0) to the point P = (x,0,z) along the solar surface (see Fig.~\ref{fig-sketch}) is found to be:
\begin{equation}
s(x) = \int_{0}^{x} dx \cdot \frac{R_{\odot}}{\sqrt{R_{\odot}^{2}-x^{2}}}
\end{equation}
leading to $s' = s/R_{\odot} = \arcsin(x')$. 

According to Eq.~(\ref{eq-B_norm}), and with $r'^{2} = x'^{2} + (z'+\lambda')^{2}$, the behaviour of the magnetic field strength on the Sun's surface is given by:
\begin{equation}
\label{eq-b_surface_hori}
b_{horizontal}^{2} = \frac{B_{horizontal}^{2}}{B_{0}^{2}} 
= \frac{\lambda'^{6}}{r'^{8}} \cdot
        [4x'^{2}+(z'+\lambda')^{2}]
\end{equation}
and
\begin{equation}
\label{eq-b_surface_verti}
b_{vertical}^{2} = \frac{B_{vertical}^{2}}{B_{0}^{2}} = \frac{\lambda'^{6}}{4r'^{8}} \cdot
        [x'^{2}+4(z'+\lambda')^{2}]
\end{equation}
for the horizontal ($\theta = 90^{\circ}$) and vertical dipole ($\theta = 0^{\circ}$), respectively. 

\begin{figure}[t]
   \includegraphics[width=0.48\textwidth]{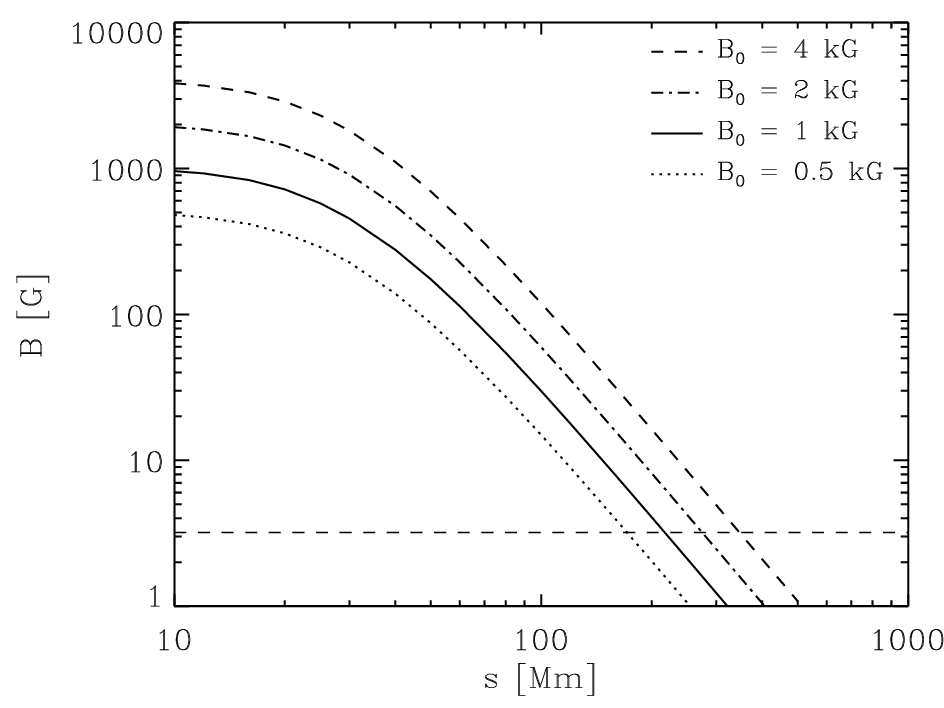}
   \caption{Behaviour of the magnetic field strength $B$ along 
           the solar surface (or photospheric level) for the horizontal dipole 
           for various values of $B_{0}$ and for $\lambda$ = 25~Mm. The horizontal dashed line indicates the quiet Sun magnetic field of 3.2~G at the photospheric level.}
\label{fig-B-surface-B0-hori}
\end{figure}

\begin{figure}[t]
   \includegraphics[width=0.48\textwidth]{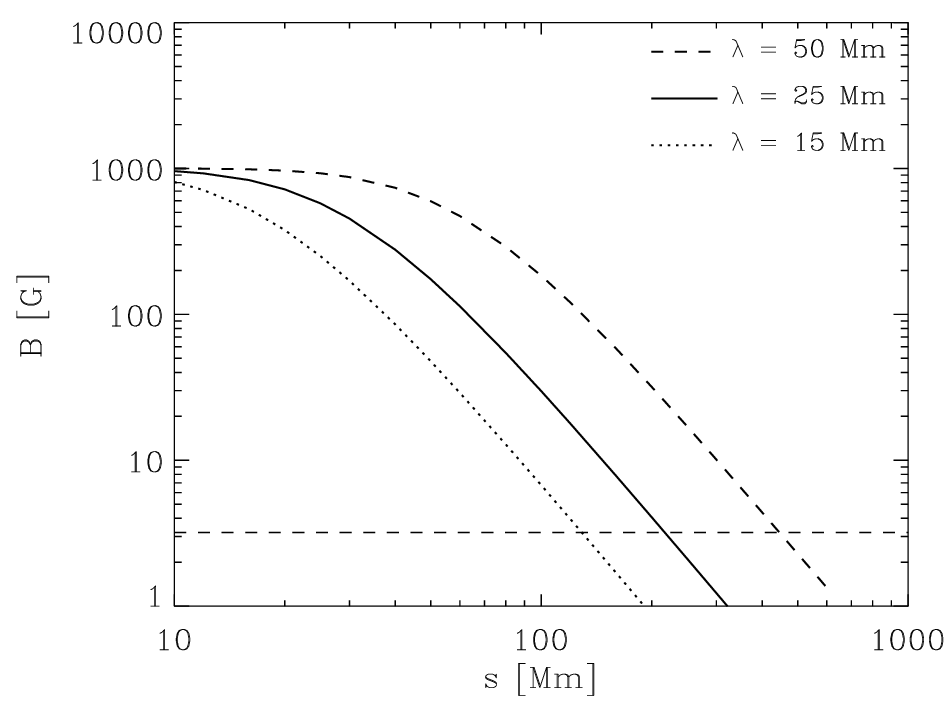}
   \caption{Behaviour of the magnetic field strength $B$ along 
           the solar surface (or photospheric level) for the horizontal dipole 
           for various values of $\lambda$ and for $B_0 = 1000$~G. The horizontal dashed line indicates the quiet Sun magnetic field of 3.2~G at the photospheric level.}            
\label{fig-B-surface-lambda-hori}
\end{figure}

\begin{figure}[t]
   \includegraphics[width=0.48\textwidth]{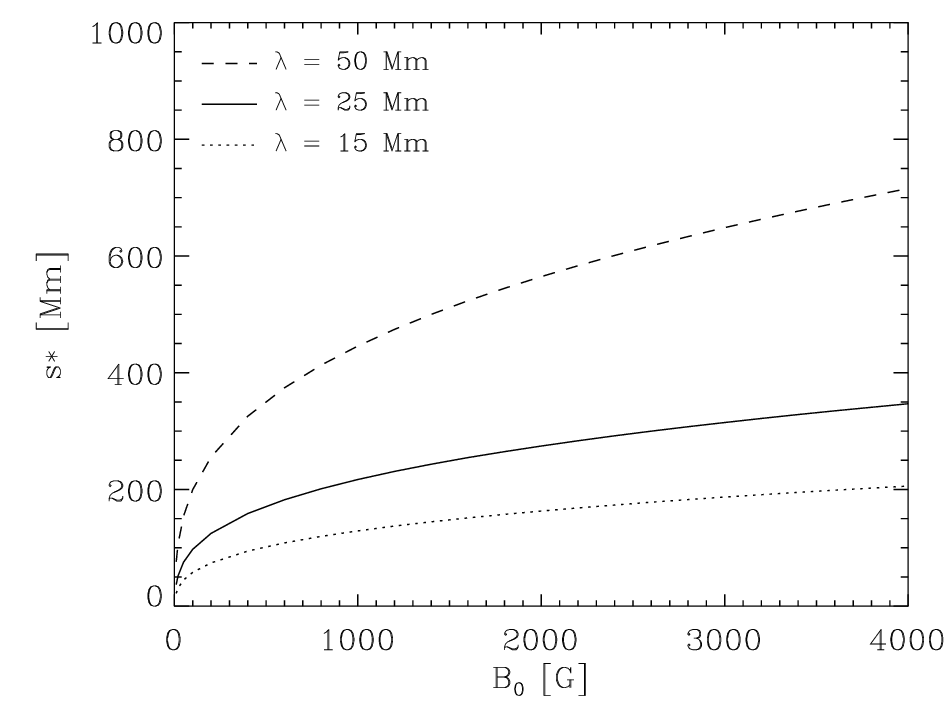}
   \caption{Dependence of the distance $s^{\ast}$ on the magnetic field $B_{0}$ 
        on the photospheric level for various values of $\lambda$ for the horizontal dipole.} 
\label{fig-B-distance-hori}
\end{figure}

We can now compute the magnetic field strength along the solar surface, i.~e. the function $b(s)$, using 
Eqs.~(\ref{eq-b_surface_hori}) and (\ref{eq-b_surface_verti}) for both cases. We present the results for various values of $B_{0}$, i.~e. $B_{0}$ = 500, 1000, 2000, 4000~G, and $\lambda$, i.~e. $\lambda$ = 15, 25, 50~Mm, in Figs.~\ref{fig-B-surface-B0-hori}, \ref{fig-B-surface-lambda-hori}, and \ref{fig-B-distance-hori}.

For the horizontal dipole, Figs.~\ref{fig-B-surface-B0-hori} and  \ref{fig-B-surface-lambda-hori}
show the decrease of the magnetic field strength with distance from the centre of the dipole (at point $P_{centre}$) on the solar surface for various values of $B_{0}$ and $\lambda$, respectively.
The influence of the dipole on its neighbourhood on the solar surface expands with increasing values of $B_{0}$ and $\lambda$. 

We define the quantity $s^{\ast}$ as the distance from the point 
P$_{centre}$ = (0,0,0) (see Fig.~\ref{fig-sketch}) to the point P on the surface of the Sun where the magnetic field strength $B$ of the dipole is equal to that of the quiet Sun, i.~e. $B(s^{\ast}) = B_{S}$. Of course, the value $B_{S}$ is not constant but can vary spatially and temporally. Here, we assume $B_{S}$ = 3.2~G (dashed lines in Figs.~\ref{fig-B-surface-B0-hori} and
\ref{fig-B-surface-lambda-hori}) as a typical value
\citep[cf.][]{Mann+1999,Klassen2000,Mann-Veronig2023}.
This value was derived from the study of extreme ultraviolet (EUV) waves in the corona
\footnote{The magnetic field of the quiet Sun cannot be measured by currently available magnetometers such as HMI and PHI/HRT, since their level
of magnetic noise is well above 3.2~G, e.g.\ in the
range 6.3$-$8.3~G for the PHI/HRT instrument (see Fig.~11 in Sihjan et al. (2022)).}.

Figure~\ref{fig-B-distance-hori} presents the dependence of $s^{\ast}$ on $B_{0}$ for the horizontal dipole. For $B_{0}$ = 1000~G and $\lambda$ = 25~Mm as typical values of active regions \citep{Stix2004,Aschwanden2005},
we find $s^{\ast}$ = 200~Mm (Fig.~\ref{fig-B-distance-hori}).
In the extreme case of $B_{0}$ = 4000~G and $\lambda$ = 50~Mm \citep[see e.~g.][]{Hale1908},
Fig.~\ref{fig-B-distance-hori} provides $s^{\ast}$ = 730~Mm, which is comparable to
the quarter of the Sun's circumference ($\pi R_{\odot}/2$ = 1100~Mm).

\begin{figure}[t]
   \includegraphics[width=0.48\textwidth]{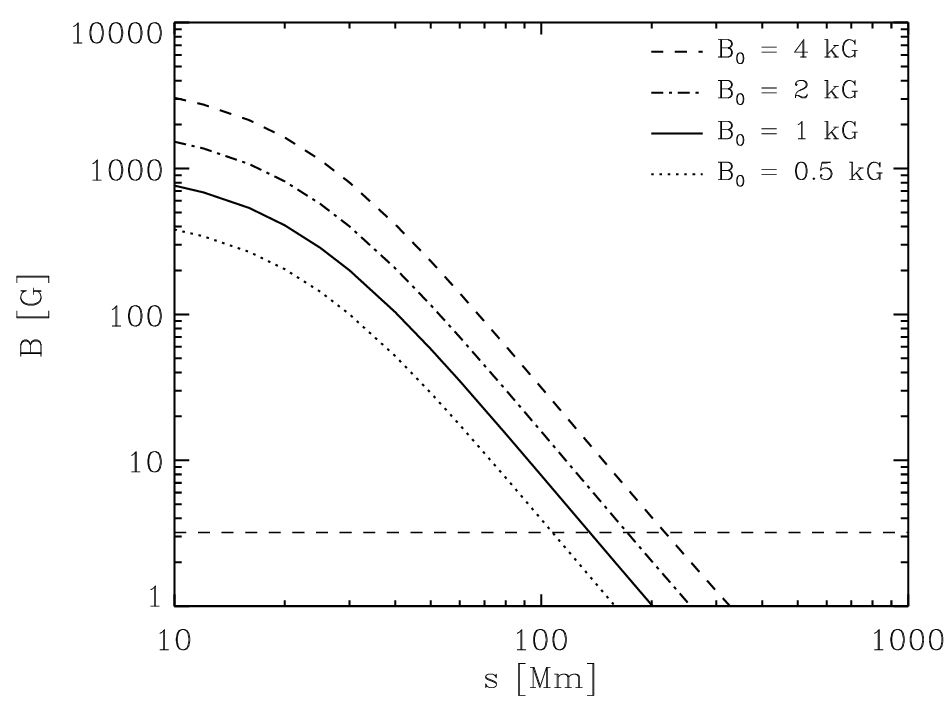}
   \caption{Behaviour of the magnetic field strength $B$ along the solar surface
   (or photospheric level) for the vertical dipole with various values of $B_{0}$ and for
   $\lambda$ = 25~Mm. The horizontal dashed line indicates the quiet Sun magnetic field
   of 3.2~G at the photospheric level.}        
\label{fig-B-surface-B0-verti}
\end{figure}

\begin{figure}[t]
   \includegraphics[width=0.48\textwidth]{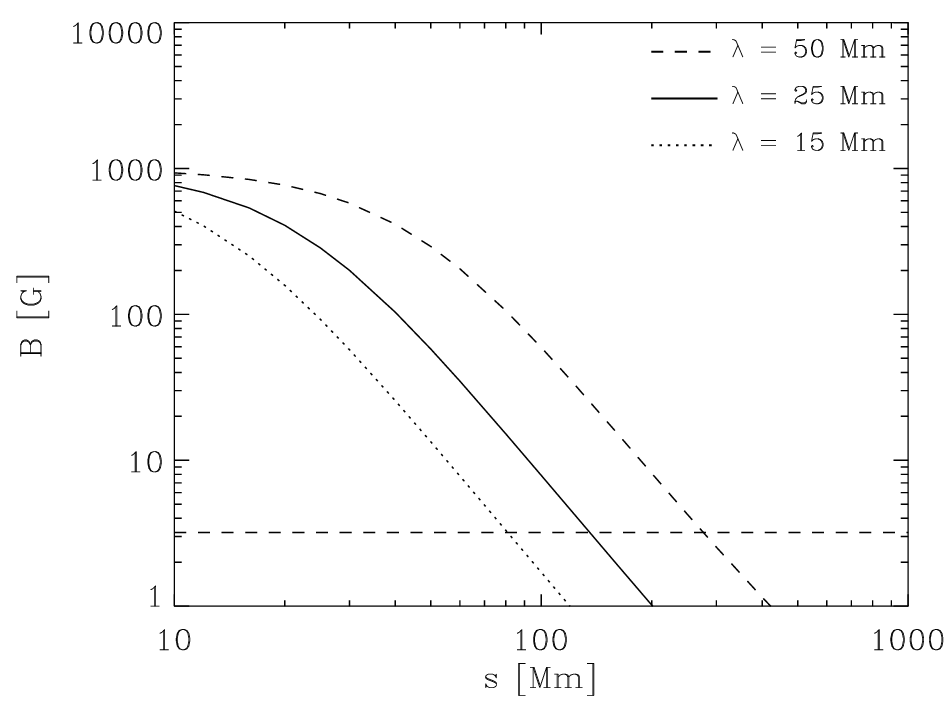}
   \caption{Behaviour of the magnetic field strength $B$ along the solar surface
   (or photospheric level) for the vertical dipole for various values of $\lambda$ and for $B_{0}$ = 1000~G. The horizontal dashed line indicates the quiet Sun magnetic field of 3.2~G at
   the photospheric level.}            
\label{fig-B-surface-lambda-verti}
\end{figure}

\begin{figure}[t]
   \includegraphics[width=0.48\textwidth]{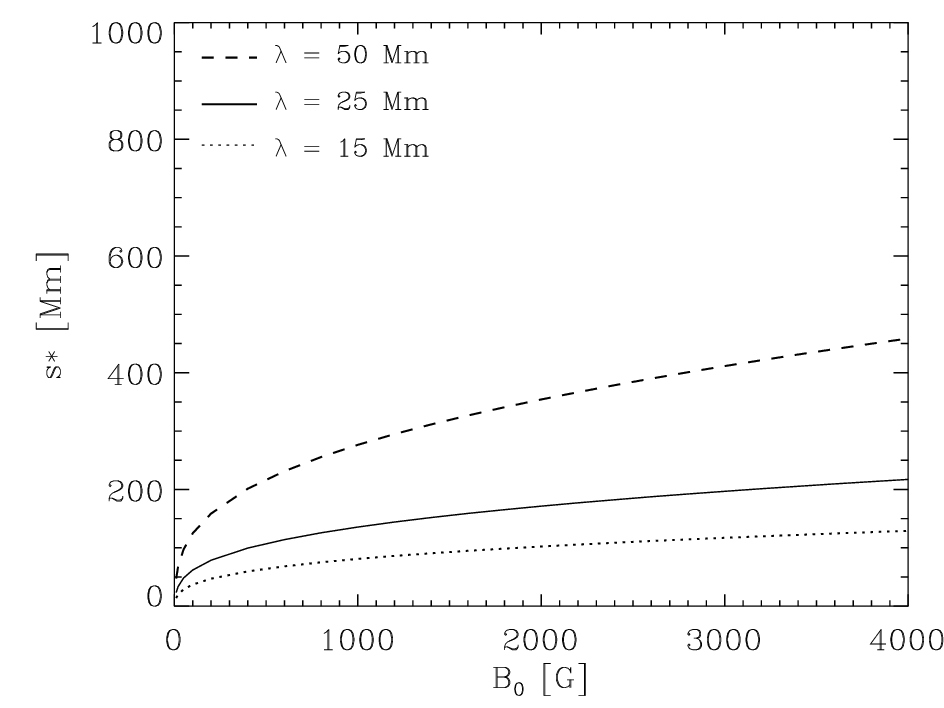}
   \caption{Dependence of the distance $s^{\ast}$ on the magnetic field $B_{0}$
   on the photospheric level for various values of $\lambda$ for the vertical dipole.} 
\label{fig-B-distance-verti}
\end{figure}

In the case of the vertical dipole, Figs.~\ref{fig-B-surface-B0-verti} and 
~\ref{fig-B-surface-lambda-verti} show the behaviour of the magnetic field strength along the distance on the solar surface from the point P$_{centre}$. The influence of the dipole expands with increasing $B_{0}$ and $\lambda$, as for the horizontal dipole.
Figure~\ref{fig-B-distance-verti} presents the dependence of $s^{\ast}$ on $B_{0}$ for various values
of $\lambda$. For $B_{0}$ = 1000~G and $\lambda$ = 25~Mm,
we derive $s^{\ast} \approx$ 130~Mm. In the extreme case of $B_{0}$ = 4000~G and $\lambda$ = 50~Mm, we find $s^{\ast}$ = 470~Mm (Fig.~\ref{fig-B-distance-verti}).

In summary, the spatial extent of the influence of the magnetic field of the dipole on its neighbourhood on the solar surface is increasing with increasing values of the parameters $B_{0}$ and $\lambda$ (see Figs.~\ref{fig-B-distance-hori} and \ref{fig-B-distance-verti}).
In addition, Figs.~\ref{fig-B-distance-hori} and \ref{fig-B-distance-verti} reveal that the influence of the magnetic field of the dipole on its neighbourhood extends further in the case of the horizontal dipole than for the vertical one (for the same values of $B_{0}$ and $\lambda$).
\subsection{\label{sec-behaviour-radial}Radial behaviour of the magnetic field above an active region}
\begin{figure}[t]
   \includegraphics[width=0.48\textwidth]{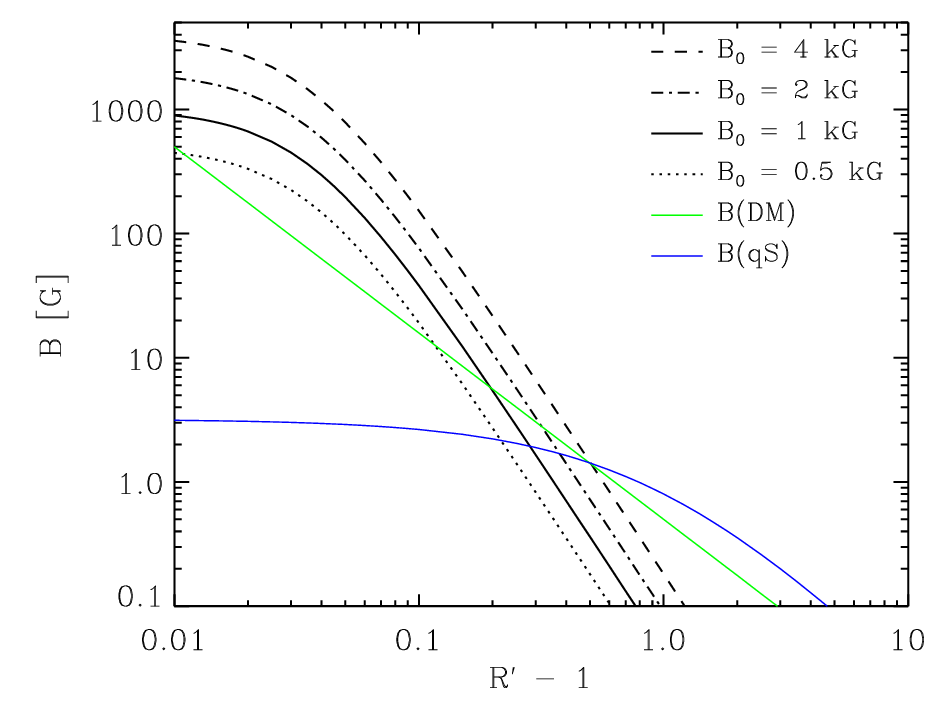}
   \caption{Radial behaviour of the magnetic field strength $B$
of a dipole for various values of $B_{0}$ for $\lambda$ = 25~Mm. 
For comparison, the radial behaviour of the magnetic field above active regions
(green line) according to\citet{Dulk1978} (see Eq.~(\ref{eq-B-DM}))
and above quiet regions (blue line) according to Eq.~(\ref{eq-B-quiet})
are also drawn.}
\label{fig-radial-B-B0}
\end{figure}

\begin{figure}[t]
   \includegraphics[width=0.48\textwidth]{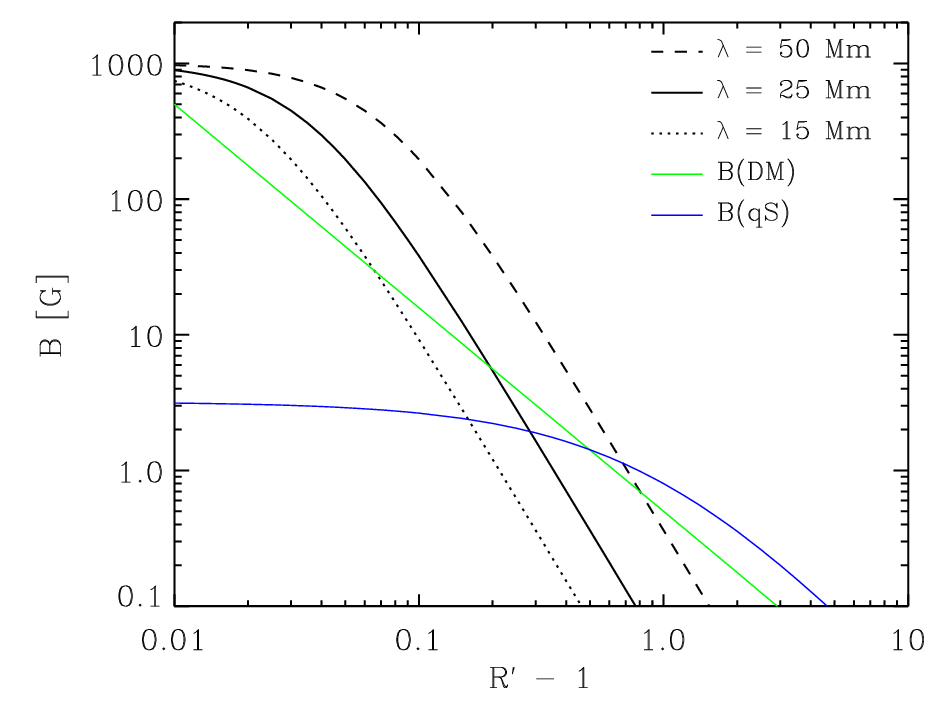}
   \caption{Radial behaviour of the magnetic field strength $B$
of a dipole for various values of $\lambda$ and for $B_{0}$ = 1000~G.
The green and blue lines are as in Fig.~\ref{fig-radial-B-B0}.}
\label{fig-radial-B-lambda}
\end{figure}

As shown in Appendix~A, the behaviour of the magnetic field strength along the z-axis is the same for both the vertical and horizontal dipoles at $x = y = 0$.
The vertical (or radial) behaviour of the magnetic field strength of a dipole is
given by Eq.~(\ref{eq-B_z_axis-hori}), which can also be written as:
\begin{equation}
\label{eq-B-dipole}
B_{dipole} = \frac{B_{0}}{\left (1 + \frac{[R-R_{\odot}]^{2}}{\lambda^{2}} \right )^{3/2}} =
\frac{B_{0}}{\left (1 + \frac{z^{2}}{\lambda^{2}} \right )^{3/2}}
\end{equation}
with $R = z + R_{\odot}$. Figures~\ref{fig-radial-B-B0} and \ref{fig-radial-B-lambda} illustrate the radial behaviour of the magnetic field of a magnetic dipole for various values of $B_{0}$ and $\lambda$. As can be seen, the radial extent of the influence of the magnetic field of the dipole increases with increasing values of $B_{0}$ and $\lambda$.

In the corona, far away from an active region, the magnetic field of the quiet Sun is given by \citep{Mann+1999,Mann+2003,Mann2015}:
\begin{equation}
\label{eq-B-quiet}
B_{qS} = B_{S} \cdot \left ( \frac{R_{\odot}}{R} \right )^{2}
\end{equation}
Then, the total magnetic field in the corona $B$ is the sum of those of the active region ($B_{ar} \, = \, B_{dipole}$) and of the quiet Sun $B_{qS}$: $B = B_{dipole} + B_{qS}$.
\footnote{Note that $B_{dipole}$ and $B_{qS}$ are actually vector-like quantities.}

For comparison, the radial behaviour of the quiet Sun (see Eq.~(\ref{eq-B-quiet})) is drawn as a blue line for $B_{S}$ = 3.2~G (see discussion above) in Figs.~\ref{fig-radial-B-B0} and \ref{fig-radial-B-lambda}. They show that the magnetic field of the quiet Sun exceeds that of the dipole at radial distances beyond the range 
$0.3 \leq (R'-1) \leq 1.0$ (with $R' = R/R_{\odot}$), depending on the values of the parameters $B_{0}$ and $\lambda$.
This result agrees well with solar observations \citep[see e.\ g.][]{Mann-Veronig2023}.
\begin{figure}[t]
   \includegraphics[width=0.48\textwidth]{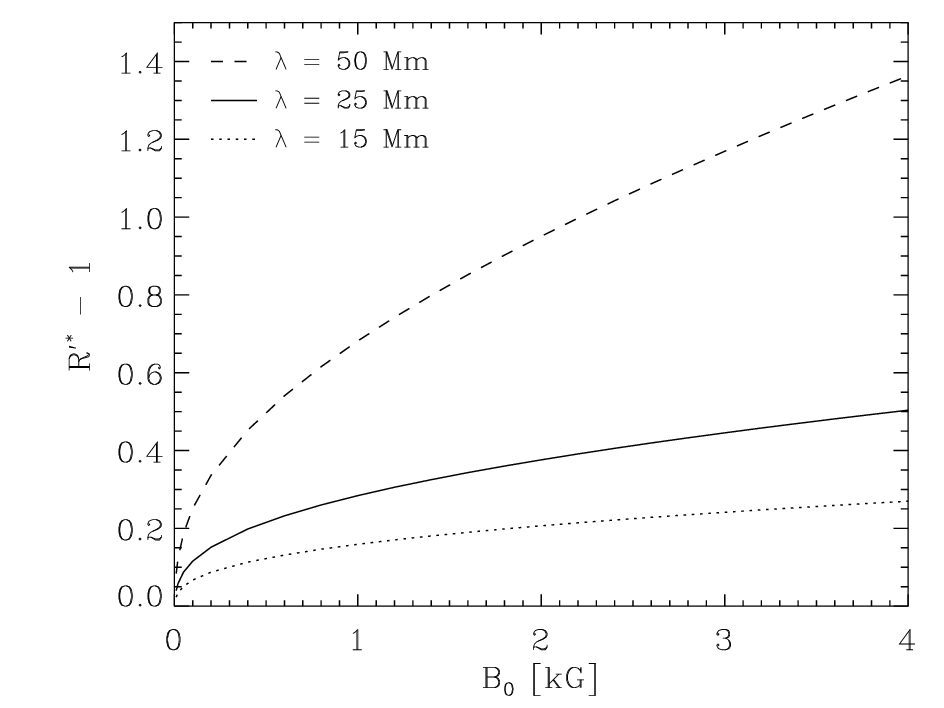}
   \caption{Radial dependence of the distance $R'^{\ast}$ on the 
            magnetic field $B_{0}$ of a dipole for various values of 
            $\lambda$.} 
\label{fig-R-prime-star}
\end{figure}
\citet{Dulk1978} derived from solar radio observations a model describing
the radial behaviour of the magnetic field above active regions in the range
$1.02 \leq R' \leq 10$. It is given by:
\begin{equation}
\label{eq-B-DM}
B_{DM} = (R'-1)^{-1.5} \times 0.5~G 
\end{equation}
This model is also drawn as a green line in Figs.~\ref{fig-radial-B-B0} and
\ref{fig-radial-B-lambda}.

Eq.~(\ref{eq-B-dipole}) contains two parameters, namely the maximum
magnetic field strength $B_{0}$ in the active region and its
spatial size $\lambda$. Hence, it describes roughly 
the vertical dependence of the magnetic field above 
an individual active region. In contradiction to that,
Eq.~(\ref{eq-B-DM}) doesn't contain any parameter and, hence, cannot
be used for individual active region. In this respect, 
our approach (see Eq.~(\ref{eq-B-dipole})) goes beyond the formula by 
\citet{Dulk1978}, i.~e. it has a broader validity.

There are also other methods to derive the vertical
dependence of the magnetic field above active regions, for instance by radio measurements
\citep{Anfi2019,Ryabov1999,Brosius2006,Bogod2016,Fleishman2010, Fleishman2020,Kuroda2020,Yu2020,Alissand2024},
and the opitcal Zeeman method \citep{Kuridze2019}.
\footnote{See also:
\url{https://www.leibniz-kis.de/en/research/solar-magnetism/coronal-magnetic-field/}}
In these papers, the vertical behaviour of the 
magnetic field is derived for different active regions,
i.~e. they are case studies. In contradiction to this
approach, our approach in terms of Eq.~(\ref{eq-B-dipole}) 
describes the vertical behaviour of the magnetic field
of any active region because of the free choice of
the parameters $B_{0}$ and $\lambda$, which are different
from case to case.

We now define the quantity $R'^{\ast} = R^{\ast}/R_{\odot}$ as the radial distance
where the magnetic field strength of the dipole is
equal to that of the quiet Sun, i.~e. $B_{dipole}(R'^{\ast}) = B_{qS}(R'^{\ast}$).
Employing Eqs.~(\ref{eq-B-dipole}) and (\ref{eq-B-quiet}), $R'^{\ast}$ can be calculated
for given values of $B_{0}$, $B_{S}$, and $\lambda$. The dependence of $R'^{\ast}$ on $B_{0}$ 
is presented for $B_{S} \, = \, 3.2$~G and for several values of $\lambda$ 
in Figure~\ref{fig-R-prime-star}.
It shows that the influence of the magnetic field of an active region extends further for larger values of $B_{0}$ and $\lambda$ in the corona.
We get  $R'^{\ast} \approx$ 1.3 for typical values $B_{0}$ = 1000~G and $\lambda$ = 25~Mm,
and $R'^{\ast} \approx$ 2.4 in the extreme case of $B_{0}$ = 4000~G and $\lambda$ = 50~Mm.
\section{Summary}\label{sec-summary}
In this paper, we have discussed the influence of the magnetic field of an active region
on its environment in the solar atmosphere and on the photosphere.
The knowledge of that is of special interest
for evaluating solar observations, e.g. for EUV waves
\citep[see e.~g.][]{Vrsnak2002,Klassen2000,Mann-Veronig2023,Warmuth2005} 
and solar radio data \citep[e.~g.][]{Aurass2005,Vocks2018,Mann+2023}.
The magnetic field of an active region is modelled by a magnetic
dipole, which is located at a depth $\lambda$ under the photosphere,
and has an angle $\vartheta$ with respect to the local vertical axis.

As discussed in Sect.~\ref{sec-influence}, the dipole horizontally directed with respect to the solar surface ($\vartheta$ = 90$^{\circ}$) models roughly a bi-polar
active region with a well developed neutral line, whereas 
the vertical dipole ($\vartheta$ = 0$^{\circ}$) describes a more unipolar active region
surrounded with diffuse areas with oppositely directed 
magnetic fields and with a circular neutral line.
Both cases are completely described by two parameters, namely $B_{0}$ and $\lambda$.

Both parameters can be derived quickly from photospheric observations 
as delivered by instruments currently in operation, such as HMI and PHI/HRT
onboard the spacecraft {\it Solar Dynamics Observatory} and {\it Solar Orbiter}, respectively.
The value of the parameter $B_{0}$ is given by the maximum of the magnetic field strength
in the active region.  
The parameter $\lambda$ can be determined by the distance between the areas of oppositely
directed magnetic fields. Then, the value of the parameter $\lambda$ is given by once
and half of this distance for the case of the horizontal and vertical dipole, respectively.
The knowledge of these parameters allows us to compute the magnetic field strength at
different places in the solar atmosphere by using the method presented in this paper. 
That demonstrates the advantage of the description of an active region by a magnetic dipole in comparison to the magnetic field extrapolation
(see \citet{Wiegelmann2021} as a review and \citet{Schrijver2003}),
which needs accurate magnetic field measurements at the photospheric level. 

With respect to the radial behaviour of the magnetic field in the corona, it should be emphasised 
that our approach goes beyond that of \citet{Dulk1978} (see Eq.~\ref{eq-B-DM}) 
with Eqs.~\ref{eq-B-dipole} and \ref{eq-B-quiet}, since it takes into account different 
values of the magnetic field in the corresponding active region.

For example, if $B_{0}$ = 800~G and $\lambda$ = 20~Mm would be  derived from photospheric measurements, a magnetic field of 6.4~G is expected at a height of 80~Mm 
above the photosphere according to Eq.~(\ref{eq-B-dipole}). There, a magnetic field of 2.6~G is expected for the quiet Sun (see Eq.~(\ref{eq-B-quiet})). The quiet Sun's magnetic field exceeds the one of the active region at a height of 140~Mm above the photosphere (see Fig.~\ref{fig-R-prime-star}).

This example demonstrates impressively that the method presented in this paper allows to estimate the spatial extent of the influence of the magnetic field of an active region on its environment in the solar atmosphere in a quick manner. But it must be emphasised once again that this is only a rough estimate, since the magnetic field topology of an active region is much more complex than that of a dipole.
A proper description of the magnetic field in the corona requires the use of modern numerical tools of magnetic field extrapolation, with which the magnetic field measured at the photospheric level can be continued into the corona 
\citep[see][for a review]{Wiegelmann2021}.
For instance, one well-known method is the potential-field-source-surface (PFSS) model \citep{Schrijver2003}. It allows to determine the topology of the magnetic field in the corona in a highly accurate way. But it requires a precisely measured photospheric magnetic field map as an input for the extrapolation.
Nevertheless, the simple approach with a magnetic dipole presented in this paper provides a quick and easy way to estimate the strength of the magnetic field in the neighbourhood of an active region. 
\section*{Acknowledgments}
The authors thank the referees for their fruitful hints
and comments, which are included in the manuscript.
\appendix
\section{The magnetic dipole\label{sec-model_a}}
The magnetic field $\vec B$ of a dipole with the moment $\vec M$ is given by
\begin{equation}
\label{eq-vector-B}
\vec B = 3 \cdot \frac{(\vec M \cdot \vec r)\vec r}{r^{5}} - \frac{\vec M}{r^{3}}
\end{equation}
at the point $\vec r$ \citep{Landau1975}. 
As presented in Figure~\ref{fig-sketch}, the framework employed in this paper is chosen in the following way: 
The z-axis is vertically directed away from the Sun. The centre of the Sun is located at the point 
$P_{\odot}$ = (0,0,-$R_{\odot}$) ($R_{\odot}$, radius of the Sun). The magnetic dipole is located 
in the x-z plane at a depth z = -$\lambda$ and takes an angle $\vartheta$ with respect to the z-axis. Then, $\vec r$ is given by $\vec r = (x,y,z+\lambda)$. The magnetic moment $\vec M = M (\sin\vartheta,0,\cos\vartheta)$ can be expressed by the magnetic field $B_{0}$ at the point $P_{centre}$ = (0,0,0) by means of Eq.~(\ref{eq-vector-B}):
\begin{equation}
\label{eq-mag-moment}
M = \frac{\lambda^{3}B_{0}}{\sqrt{1+3\cos^{2}\vartheta}}
\end{equation}
Inserting Eq.~(\ref{eq-mag-moment}) into Eq.~(\ref{eq-vector-B}),
the components of the magnetic field of the dipole can be written as 
\begin{eqnarray}
\label{eq-B_x}
\frac{B_{x}}{B_{0}} & = & \frac{\lambda^{3}}{\sqrt{1+3\cos^{2}\vartheta}} \cdot \frac{1}{r^{5}} \nonumber \\ 
                    & &\times \left\{ 3[x\sin\vartheta + (z+\lambda)\cos\vartheta]x - r^{2}\sin\vartheta \right\}
\end{eqnarray}
\begin{eqnarray}
\label{eq-B_y}
\frac{B_{y}}{B_{0}} & = & \frac{\lambda^{3}}{\sqrt{1+3\cos^{2}\vartheta}} \cdot \frac{1}{r^{5}} \nonumber \\ 
                    & &\times \left\{ 3[x\sin\vartheta + (z+\lambda)\cos\vartheta]y \right\}
\end{eqnarray}
and
\begin{eqnarray}
\label{eq-B_z}
\frac{B_{z}}{B_{0}} & = & \frac{\lambda^{3}}{\sqrt{1+3\cos^{2}\vartheta}} \cdot \frac{1}{r^{5}} \nonumber \\ 
                    & & \left\{ 3[x\sin\vartheta + (z+\lambda)\cos\vartheta](z+\lambda) - r^{2}\cos\vartheta \right\}
\end{eqnarray}
in Cartesian coordinates. Then, the magnetic field strength $B = (B_{x}^{2}+B_{y}^{2}+B_{z}^{2})^{1/2}$ 
is found to be:
\begin{eqnarray}
\label{eq-B_norm}
\frac{B}{B_{0}} & = & \frac{\lambda^{3}}{\sqrt{1+3\cos^{2}\vartheta}} \cdot \frac{1}{r^{4}} \nonumber \\ 
                    & & \times \sqrt{3[x\sin\vartheta + (z+\lambda)\cos\vartheta]^{2} + r^{2}}
\end{eqnarray}
with $r = [x^{2}+y^{2}+(z+\lambda)^{2}]^{1/2}$.

For illustration, the behaviour of the magnetic field is investigated for the horizontal (i.~e. $\vartheta = 90^{\circ}$) and vertical (i.~e. $\vartheta = 0^{\circ}$) dipole.

\begin{figure}[t]
\includegraphics[width=0.48\textwidth]{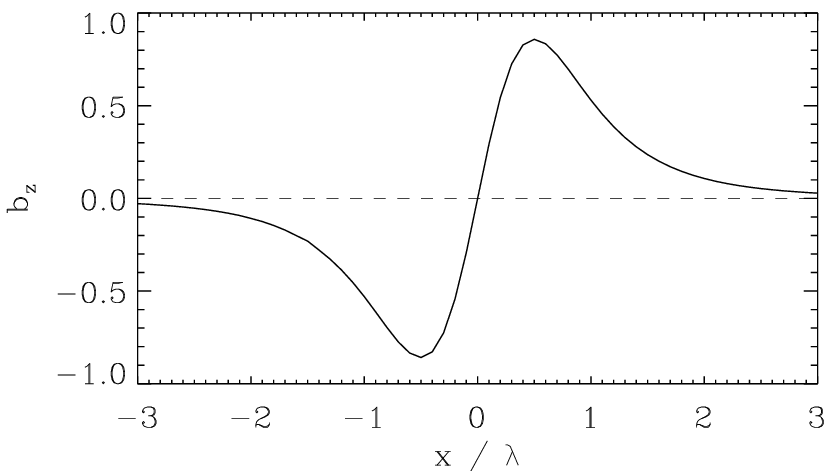}
\caption{Behaviour of the z-component (see Eq.~(\ref{eq-B_z_hori}))
of the magnetic field along the x-axis at y = z = 0 for the horizontal dipole.}   
\label{fig-B-hori}
\end{figure}

In the case of the horizontal dipole, the behaviour of z-component of the magnetic field is studied along the x-axis for $y = z = 0$. For this case, the Eq.~(\ref{eq-B_z}) is reduced to:
\begin{equation}
\label{eq-B_z_hori}
b_{z} = \frac{B_{z}}{B_{0}} = \frac{3\eta}{(1+\eta^{2})^{5/2}}
\end{equation}
with $\eta = x/\lambda$. The spatial behaviour of the z-component $B_{z}$ of the magnetic field is shown in Figure~\ref{fig-B-hori}.
The inspection of Eq.~(\ref{eq-B_z_hori}) and Figure~\ref{fig-B-hori}
reveals that the z-component of the magnetic field is an odd function with respect to $\eta$ with a zero at $\eta = 0$, a local minimum at $\eta = -1/2$ with $b_{z,min} = b_{z}(\eta = -1/2) = -0.8587$, and a local maximum at $\eta = 1/2$ with $b_{z,max} = b_{z}(\eta = 1/2) = 0.8587$.
Along the z-axis for $x = y = 0$, the spatial behaviour of the magnetic field strength is given by:
\begin{equation}
\label{eq-B_z_axis-hori}
b = \frac{B}{B{_0}} = -b_{x} = \frac{1}{(1+\xi^{2})^{3/2}}
\end{equation}
with $b_{x} = B_{x}/B_{0}$ and $\xi = z/\lambda$ (see Eq~(\ref{eq-B_norm})). 

\begin{figure}[t]
\includegraphics[width=0.48\textwidth]{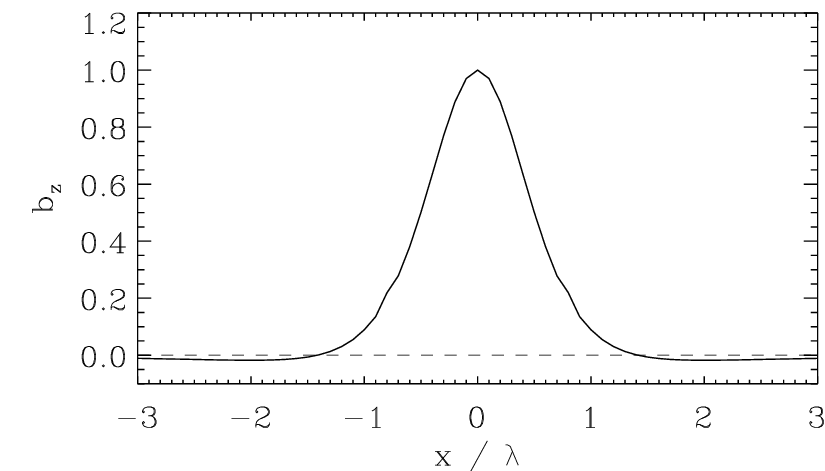}
\caption{Behaviour of the z-component (see Eq.~(\ref{eq-B_z_verti})
of the magnetic field along the x-axis at y = z = 0 for the vertical dipole.}
\label{fig-B-verti}
\end{figure}

In the case of the vertical dipole, the behaviour of the z-component of the magnetic field along the x-axis for $y = z = 0$ is given by 
\begin{equation}
\label{eq-B_z_verti}
b_{z} = \frac{B_{z}}{B_{0}} = \frac{1}{2} \cdot \frac{(2-\eta^{2})}{(1+\eta^{2})^{5/2}}
\end{equation}
(see Eq.~(\ref{eq-B_z}). The spatial behaviour of the z-component $B_{z}$ of the magnetic field is depicted in Figure~\ref{fig-B-verti}.
The inspection of Eq.~(\ref{eq-B_z_verti}) and Figure~\ref{fig-B-verti}
reveal that the z-component of the magnetic field is an even function with respect to $\eta$ with a local maximum at $\eta = 0$ with $b_{z}(\eta = 0) = 1$ and two minima at $\eta = \pm 2.0$, with $b_{z,min} = b_{z}(\eta = \pm$2) = -0.0179. Along the vertical axis, the spatial behaviour of the magnetic field strength is given by:
\begin{equation}
\label{eq-B_z_axis-verti}
b = b_{z} = \frac{1}{(1+\xi^{2})^{3/2}}
\end{equation}
with $\xi = z/\lambda$ and for $x=y=0$ (see Eq.~(\ref{eq-B_norm})).
Note that the behaviour of the magnetic field strength along the z-axis is the same for both the vertical and horizontal dipole at $x = y = 0$ (compare Eqs.~(\ref{eq-B_z_axis-verti}) and 
(\ref{eq-B_z_axis-hori})).

In summary, the magnetic field of a dipole is completely determined only by three parameters: $B_{0}$, $\vartheta$, and $\lambda$. 

\bibliography{AN_MAG}

\end{document}